# Fastest Mixing Markov Chain on Symmetric $K$-Partite Network

Saber Jafarizadeh
Department of Electrical Engineering
Sharif University of Technology, Azadi Ave, Tehran, Iran
Email: jafarizadeh@ee.sharif.edu

*Abstract*—Solving fastest mixing Markov chain problem (i.e. finding transition probabilities on the edges to minimize the second largest eigenvalue modulus of the transition probability matrix) over networks with different topologies is one of the primary areas of research in the context of computer science and one of the well known networks in this issue is $K$-partite network. Here in this work we present analytical solution for the problem of fastest mixing Markov chain by means of stratification and semidefinite programming, for four particular types of $K$-partite networks, namely Symmetric $K$-PPDR, Semi Symmetric $K$-PPDR, Cycle $K$-PPDR and Semi Cycle $K$-PPDR networks.

Our method in this paper is based on convexity of fastest mixing Markov chain problem, and inductive comparing of the characteristic polynomials initiated by slackness conditions in order to find the optimal transition probabilities.

The presented results shows that a Symmetric $K$-PPDR network and its equivalent Semi Symmetric $K$-PPDR network have the same *SLEM* despite the fact that Semi symmetric $K$-PPDR network has less edges than its equivalent symmetric $K$-PPDR network and at the same time symmetric $K$-PPDR network has better mixing rate per step than its equivalent semi symmetric $K$-PPDR network at first few iterations. The same results are true for Cycle $K$-PPDR and Semi Cycle $K$-PPDR networks. Also the obtained optimal transition probabilities have been compared with the transition probabilities obtained from Metropolis-Hasting method by comparing mixing time improvements numerically.

*Index Terms*— **Fastest mixing Markov chain, Sensor networks, Weight optimization, Second largest eigenvalue modulus, Semidefinite programming, Graph theory.**

## I. INTRODUCTION

Markov chain problem is a well studied topic with a wide range of applications in different fields, for instance Markov chain Monte Carlo simulation is a powerful algorithm in statistics, physics, chemistry, biology, computer science and many others (see, e.g., [1, 2]). The main application of Markov chain simulation is to the random sampling of a huge state space with a specified probability distribution. The basic idea is to construct a Markov chain that converges asymptotically to the specified equilibrium distribution. The efficiency of such algorithms depends on how fast the constructed Markov chain converges to the equilibrium, i.e., how fast the chain mixes. Mixing rate of Markov chains are determined by their Second Largest Eigenvalue Modulus (*SLEM*) and the problem of finding the Markov chains with maximum possible mixing rate is well known as Fastest Mixing Markov Chain (FMMC) problem. FMMC problem in the context of computer science has found many practical applications including fast load balancing of parallel computing systems [3, 4], and in average consensus and gossip algorithms in sensor networks [5, 6]. Most previous works are focused on bounding the SLEM of a Markov chain with various numerical techniques, and developing some heuristics to assign transition probabilities to obtain faster mixing Markov chains. Boyd et al. [7] have shown that FMMC problem can be formulated as a convex optimization problem, in particular a semidefinite program and in [8] the FMMC problem have been solved for a path network based on conjectured optimal transition probabilities by [9]. Recently in [10, 11] the author has solved fastest distributed consensus averaging problem analytically for complete cored, symmetric and asymmetric star networks and in [12, 13] for tree and chain of rhombus networks by means of stratification and semidefinite programming.

One of basic and common types of networks is bipartite network. Bipartite graphs find applications in modern coding theory. Two famous examples of bipartite graphs are factor graphs and tanner graphs where factor graphs are widely used for decoding LDPC and turbo codes and tanner graphs are used to construct longer codes from smaller ones, also bipartite graphs are useful for modeling matching problems. A more generalized form of bipartite graphs is *K*-partite graphs. *K*-partite graphs are extensively used in the context of computer science, particularly data mining and unsupervised learning. Various data mining applications involve data objects of multiple types that are related to each other, which can be naturally formulated as a *K*-partite graph. For example, Web pages, search queries, and Web users in a Web search system form a tri-partite graph.

Here in this work, we have provided analytical solution for FMMC problem over four particular types of *K*-partite networks, namely symmetric *K*-PPDR, semi symmetric *K*-PPDR, cycle *K*-PPDR and semi cycle *K*-PPDR, by means of stratification and semidefinite programming. Our method in this paper is based on convexity of FMMC problem, and inductive comparing of the characteristic polynomials initiated by slackness conditions in order to find the optimal probabilities.

Some numerical simulations are carried out to investigate the trade-off between a symmetric $K$-PPDR network and its equivalent semi symmetric $K$-PPDR network and it has been shown that a symmetric $K$-PPDR network and its equivalent semi symmetric $K$-PPDR network have the same *SLEM* despite the fact that semi symmetric $K$-PPDR network has less edges than its equivalent symmetric $K$-PPDR network and at the same time symmetric $K$-PPDR network has better mixing rate per step than its equivalent semi symmetric $K$-PPDR network at first iterations. The same results are true for cycle $K$-PPDR and semi cycle $K$-PPDR networks. Also the obtained optimal transition probabilities have been compared with the transition probabilities obtained from Metropolis-Hasting method by comparing mixing time improvements numerically.

The organization of the paper is as follows. Section II is an overview of the materials used in the development of the paper, including relevant concepts from Fastest Mixing Markov Chain, graph symmetry and semidefinite programming. Section III contains the main results of the paper where symmetric $K$-PPDR, semi symmetric $K$-PPDR, cycle $K$-PPDR and semi cycle $K$-PPDR are introduced together with the corresponding evaluated *SLEM* and obtained optimal probabilities. Section IV presents simulations, demonstrating improvements of the obtained optimal transition probabilities over other transition probabilities obtained from Metropolis-Hasting method by analyzing *SLEM* and time to convergence along with comparison of symmetric $K$-PPDR and semi symmetric $K$-PPDR networks in a per step manner. Section V is devoted to proof of main results of paper for symmetric $K$-PPDR network and section VI concludes the paper.

## II. Preliminaries

This section introduces the notation used in the paper and reviews relevant concepts from fastest mixing Markov chain problem, graph symmetry and semidefinite programming.

### A. Distributed Consensus

We consider a network $\mathcal{N}$ with the associated graph $\mathcal{G} = (\mathcal{V}, \mathcal{E})$ consisting of a set of nodes $\mathcal{V}$ and a set of edges $\mathcal{E}$ where each edge $\{i, j\} \in \mathcal{E}$ is an unordered pair of distinct nodes.

We define a discrete-time Markov chain by associating with each edge $\{i, j\} \in \mathcal{E}$ a transition probability $p_{i,j}$ ($p_{i,i}$ denotes the holding probability at vertex $i$). We assume the transition between two vertices connected by an edge is symmetric, i.e., $p_{i,j} = p_{j,i}$. Thus the transition probability matrix, $P \in \mathbf{R}^{\mathcal{N} \times \mathcal{N}}$, satisfies $P = P^T, P \geq 0, P\mathbf{1} = \mathbf{1}$, where $P^T$ denotes the transpose of $P$, the inequality $P \geq 0$ means elementwise, and $\mathbf{1}$ denotes the column vector of all ones.

Since $P$ is symmetric and stochastic, the uniform distribution $(1/N)\mathbf{1}^T$ is stationary. In addition, the eigenvalues of $P$ are real, and no more than one in magnitude. We list them in decreasing order as

$$1 = \lambda_1(P) \geq \lambda_2(P) \geq \cdots \geq \lambda_N(P) \geq -1$$

We denote the second largest modulus (*SLEM*) of $P$ by $\mu(P)$, i.e.

$$\mu(P) = \max_{i=2,\ldots,N} |\lambda_i(P)| = \max\{\lambda_2(P), -\lambda_N(P)\}$$

In general the smaller $\mu(P)$ is, the faster the Markov chain converges to its stationary distribution.

The fastest mixing Markov chain (FMMC) problem is to find the optimal $P$ that minimizes $\mu(P)$, thus FMMC problem can be expressed as the following optimization problem:

$$\min_{P} \quad \mu(P)$$
$$s.t. \quad P = P^T, P \geq 0, P\mathbf{1} = \mathbf{1}, \quad (1)$$
$$\forall \{i,j\} \notin \mathcal{E}: P_{ij} = 0$$

In [7] it has been shown that (1) is a convex optimization problem and can be derived in the semidefinite programming form as:

$$\min \quad s$$
$$s.t. \quad -sI \preccurlyeq P - \mathbf{1}\mathbf{1}^T/n \preccurlyeq sI$$
$$P = P^T, \quad P \geq 0, \quad P\mathbf{1} = \mathbf{1} \quad (2)$$
$$\forall \{i,j\} \notin \mathcal{E}: P_{ij} = 0$$

Here $I$ denotes the identity matrix, and the variables are the matrix $P$ and the scalar $s$. The expression $Y \preccurlyeq X$ means $X - Y$ is positive semidefinite. We refer to problem (2) as the Fastest Mixing Markov Chain (FMMC) problem.

B. *Symmetry of Graphs*

An automorphism of a graph $\mathcal{G} = (\mathcal{V}, \mathcal{E})$ is a permutation $\sigma$ of $\mathcal{V}$ such that $\{i,j\} \in \mathcal{E}$ if and only if $\{\sigma(i), \sigma(j)\} \in \mathcal{E}$, the set of all such permutations, with composition as the group operation, is called the automorphism group of the graph and denoted by $Aut(\mathcal{G})$. For a vertex $i \in \mathcal{V}$, the set of all images $\sigma(i)$, as $\sigma$ varies through a subgroup $G \subseteq Aut(\mathcal{G})$, is called the orbit of $i$ under the action of $G$. The vertex set $\mathcal{V}$ can be written as disjoint union of distinct orbits. In [14], it has been shown that the transition probabilities on the edges within an orbit must be the same.

C. *Semidefinite Programming*

SDP is a particular type of convex optimization problem [15]. An SDP problem requires minimizing a linear function subject to a linear matrix inequality (LMI) constraint [16]:

$$\min \quad \rho = c^T x,$$

$$s.t. \quad F(x) \geq 0$$

where $c$ is a given vector, $x^T = (x_1, \ldots, x_n)$, and $F(x) = F_0 + \sum_i x_i F_i$, for some fixed Hermitian matrices $F_i$. The inequality sign in $F(x) \geq 0$ means that $F(x)$ is positive semidefinite.

This problem is called the primal problem. Vectors $x$ whose components are the variables of the problem and satisfy the constraint $F(x) \geq 0$ are called primal feasible points, and if they satisfy $F(x) \geq 0$, they are called strictly feasible points. The minimal objective value $c^T x$ is by convention denoted by $\rho^*$ and is called the primal optimal value.

Due to the convexity of the set of feasible points, SDP has a nice duality structure, with the associated dual program being:

$$\max \quad -Tr[F_0 Z]$$

$$s.t. \quad Z \geq 0$$

$$Tr[F_i Z] = c_i$$

Here the variable is the real symmetric (or Hermitian) positive matrix $Z$, and the data $c$, $F_i$ are the same as in the primal problem. Correspondingly, matrix $Z$ satisfying the constraints is called dual feasible (or strictly dual feasible if $Z > 0$). The maximal objective value of $-Tr[F_0 Z]$, i.e. the dual optimal value is denoted by $d^*$.

The objective value of a primal (dual) feasible point is an upper (lower) bound on $\rho^*(d^*)$. The main reason why one is interested in the dual problem is that one can prove that $d^* \leq \rho^*$, and under relatively mild assumptions, we can have $\rho^* = d^*$. If the equality holds, one can prove the following optimality condition on $x$.

A primal feasible $x$ and a dual feasible $Z$ are optimal, which is denoted by $\hat{x}$ and $\hat{Z}$, if and only if

$$F(\hat{x})\hat{Z} = \hat{Z}F(\hat{x}) = 0. \tag{3}$$

This latter condition is called the complementary slackness condition.

In one way or another, numerical methods for solving SDP problems always exploit the inequality $d \leq d^* \leq \rho^* \leq \rho$, where $d$ and $\rho$ are the objective values for any dual feasible point and primal feasible point, respectively. The difference

$$\rho^* - d^* = c^T x + Tr[F_0 Z] = Tr[F(x)Z] \geq 0$$

is called the duality gap. If the equality $d^* = \rho^*$ holds, i.e. the optimal duality gap is zero, and then we say that strong duality holds.

## III. MAIN RESULTS

This section presents the main results of the paper. Here we have introduced $K$-Partite Pseudo Distance Regular ($K$-PPDR) network with the corresponding evaluated *SLEM* and optimal transition probabilities, proofs and more detailed discussion are deferred to Sections V.

### A. Symmetric K-PPDR Network

A $K$-partite graph is a graph whose graph vertices can be partitioned into $K$ disjoint sets so that no two vertices within the same set are adjacent. If we arrange disjoint sets in a $K$-partite graph as $S_1, \ldots, S_K$, a $K$-PPDR graph can be defined as a $K$-partite graph where nodes in each set $S_i$ are connected to all of the nodes in neighboring sets $S_{i-1}, S_{i+1}$ except the first and last sets $S_1$ and $S_K$ which are only connected to the nodes of $S_2$ and $S_{K-1}$ respectively. Also we define $n_i$ for $i = 1, \ldots, K$ as the number of nodes on $i$-th set. A symmetric $K$-PPDR network is a $K$-PPDR network where $n_1 = \cdots = n_K = n$ (see Fig.1. for $K = 6, n = 3$).

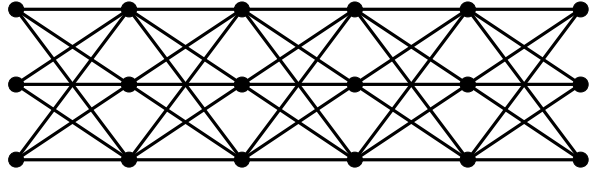

Fig.1. A symmetric $K$-PPDR network for $K = 6, n = 3$.

In section V we have proved that the optimal transition probabilities for the edges of a symmetric $K$-PPDR network equals

$$p = \frac{1}{2n} \quad \text{for } K \geq 3 \tag{4-1}$$

$$p = \frac{2}{3n} \quad \text{for } K = 2 \tag{4-2}$$

and for the *SLEM* of the network we have

$$SLEM = \cos(\pi/K) \quad \text{for } K \geq 3$$

$$SLEM = \frac{1}{3} \quad \text{for } K = 2$$

A very obvious example for symmetric $K$-PPDR network is path network which is a symmetric network with $n = 1$, where the results thus obtained are in agreement with those of [8,17].

### B. Semi Symmetric K-PPDR Network

Semi Symmetric $K$-PPDR network is a symmetric $K$-PPDR network with path graphs $P_2$ within where these path graphs are not connected to each other consecutively (see Fig.2. for $K = 6, n = 3$). In a Semi Symmetric $K$-PPDR graph there are two kinds of connectivity between the nodes of two neighboring sets, namely full and strait connectivity. In full connectivity each node of two neighboring sets is connected to all of the nodes of other set and in strait connectivity each node of two neighboring sets is connected to only one of the nodes of other set, thus the edges of a Semi Symmetric $K$-PPDR graph can be divided into two groups, namely, full edges and strait edges.

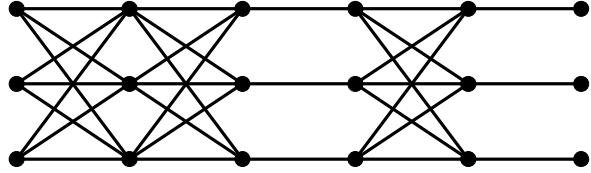

Fig.2. A Semi Symmetric $K$-PPDR network for $K = 6, n = 3$.

Using the same procedure as in section V, namely stratification and semidefinite programming we can state that on a semi symmetric $K$-PPDR network the optimal transition probabilities on the full edges equals

$$p = \frac{1}{2n} \quad \text{for } K \geq 4 \tag{5-1}$$

$$p = \frac{2}{3n} \quad \text{for } K = 3 \tag{5-2}$$

and on strait edges equals

$$p = \frac{1}{2} \tag{5-3}$$

and for the *SLEM* of the network we have

$$SLEM = \cos(\pi/K) \quad \text{for } K \geq 4$$

$$SLEM = \frac{1 + \sqrt{13}}{6} \quad \text{for } K = 3$$

C. *Cycle* K-*PPDR Network*

Cycle $K$-PPDR network is a symmetric $K$-PPDR network where the nodes of first and last sets are connected to each other as well as other neighboring sets to form a cycle of $K$-partite graphs (see Fig.3. for $K = 8, n = 2$).

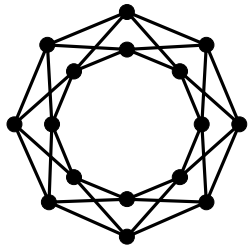

Fig.3. A Cycle $K$-PPDR network for $K = 8, n = 2$.

Automorphism of Cycle $K$-PPDR graph is $S_n$ permutation of nodes of same set, hence according to subsection II-B it has $K$ class of edge orbits and considering the symmetry of network it suffices to consider just one transition probability and consequently the Transition matrix for the network can be defined as

$$P = \begin{bmatrix} 1-2np & np & 0 & & & \\ np & 1-2np & np & 0 & & \\ 0 & np & 1-2np & \ddots & & 0 \\ & 0 & \ddots & \ddots & & np \\ & & & & np & 1-2np \end{bmatrix}$$

and its Laplacian matrix is

$$L = \begin{bmatrix} 2 & -1 & 0 & & \\ -1 & 2 & -1 & 0 & \\ 0 & -1 & 2 & \ddots & 0 \\ & 0 & \ddots & \ddots & -1 \\ & & & -1 & 2 \end{bmatrix}$$

which has eigenvalues $2 - 2\cos(2i\pi/n)$, $i = 1, \ldots, K$. Thus the second largest and smallest eigenvalues of transition matrix are

$$\lambda_2(P) = 1 - np\left(2 - 2\cos\left(\frac{2\pi}{K}\right)\right) \qquad \lambda_K(P) = 1 - np\left(2 - 2\cos\left(\frac{2\lfloor K/2 \rfloor \pi}{K}\right)\right)$$

where $\lfloor K/2 \rfloor$ denotes the largest integer that is no larger than $K/2$. Considering the second largest and smallest eigenvalues of transition matrix given above for the optimal transition probability and *SLEM* of network we have

$$p = \frac{1}{2n - n\left(\cos\left(\frac{2\pi}{K}\right) + \cos\left(\frac{2\lfloor K/2 \rfloor \pi}{K}\right)\right)}$$

and

$$SLEM = \frac{\cos\left(\frac{2\pi}{K}\right) - \cos\left(\frac{2\lfloor K/2 \rfloor \pi}{K}\right)}{2 - \cos\left(\frac{2\pi}{K}\right) - \cos\left(\frac{2\lfloor K/2 \rfloor \pi}{K}\right)}$$

D. *Semi Cycle* K-*PPDR Network*

semi Cycle $K$-PPDR network is a semi symmetric $K$-PPDR network where the nodes of first and last sets are connected to each other as well as other neighboring sets to form a cycle of $K$-partite graphs (see Fig.4. for $K = 8, n = 2$). Similar to semi symmetric $K$-PPDR network, edges of a semi Cycle $K$-PPDR network can be divided into two groups, namely, full edges and strait edges.

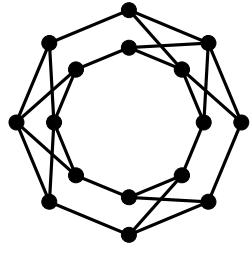

Fig.4. A Semi Cycle $K$-PPDR network for $K = 8, n = 2$.

Using the same procedure as done for Cycle $K$-PPDR in previous subsection we can state that on a semi cycle $K$-PPDR network the optimal transition probabilities on the strait edges and full edges are

$$p = \frac{1}{2} \tag{6-1}$$

and

$$p = \frac{1}{2n - n\left(\cos\left(\frac{2\pi}{K}\right) + \cos\left(\frac{2\lfloor K/2 \rfloor \pi}{K}\right)\right)} \tag{6-2}$$

respectively and for the *SLEM* of the network we have

$$SLEM = \frac{\cos\left(\frac{2\pi}{K}\right) - \cos\left(\frac{2\lfloor K/2 \rfloor \pi}{K}\right)}{2 - \cos\left(\frac{2\pi}{K}\right) - \cos\left(\frac{2\lfloor K/2 \rfloor \pi}{K}\right)}$$

## IV. PREVIEW AND DISCUSSIONS

Consider a symmetric $K$-PPDR network and semi symmetric $K$-PPDR network both having $K$ sets of $n$ disjoint nodes. Both of these networks for the choice of optimal transition probabilities given in (4) and (5) has the same *SLEM* which is $\cos(\pi/K)$ for $K \geq 4$ but the semi symmetric $K$-PPDR network has less edges, therefore a symmetric $K$-PPDR network can be replaced by its equivalent semi symmetric $K$-PPDR network while having the same *SLEM* and convergence rate by less edges and connections if and only if the transition probabilities on edges of the equivalent semi symmetric $K$-PPDR network is chosen as the optimal transition probabilities given in (5). The same relation holds true for cycle $K$-PPDR network and semi cucle $K$-PPDR network, or in other words a cycle $K$-PPDR network can be replaced by its equivalent semi cycle $K$-PPDR network while having the same *SLEM* and convergence rate by less edges and connections if and only if the transition probabilities on edges of the equivalent semi cycle $K$-PPDR network is chosen as the optimal transition probabilities given in (6).

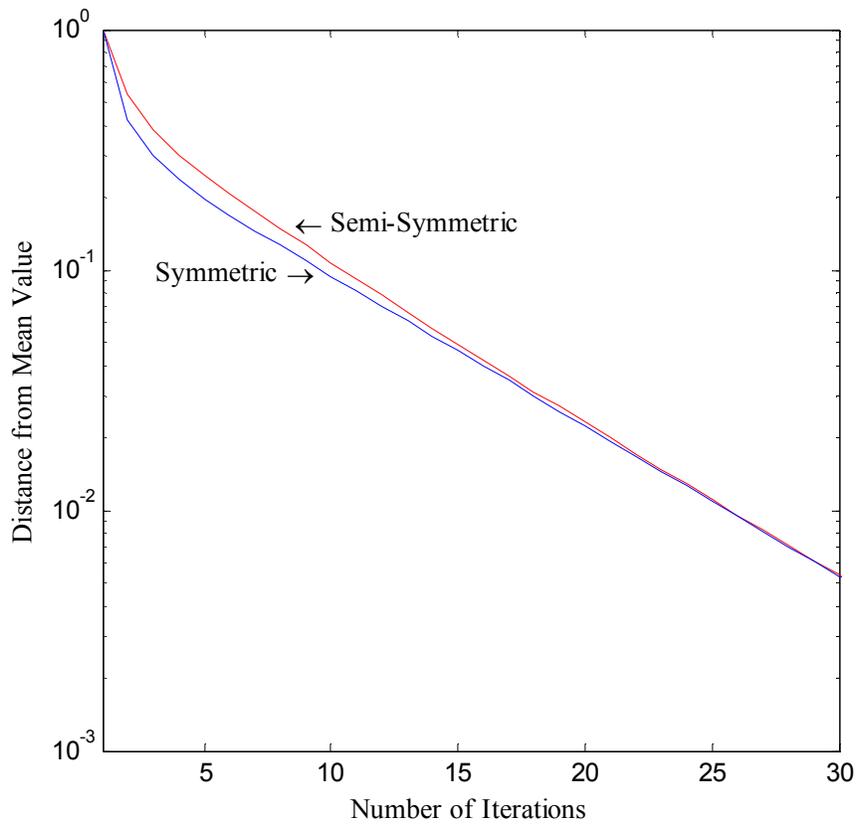

Fig.5. Normalized Euclidean Distance of vector of node values from the stationary distribution in terms of number of iterations.

In Fig.5. Normalized Euclidean distance of vector of node values from the stationary distribution in terms of number of iterations for a symmetric $K$-PPDR network and its equivalent semi symmetric $K$-PPDR network depicted in Fig.1. and Fig.2., respectively is presented. As it is obvious from Fig.5. at first iterations symmetric $K$-PPDR network has better mixing rate per step than its equivalent semi symmetric $K$-PPDR network due to its additional connections but after first few iterations both networks achieve the same mixing rate per step since they have the same *SLEM* value. Also the results depicted in Fig.5. implies that *SLEM* is an asymptotic convergence factor.

Another important issue is that in a symmetric $K$-PPDR network the obtained optimal transition probabilities and the transition probabilities obtained from Metropolis-Hasting algorithm are the same which does not hold true for other three kinds of networks studied in this paper. To show optimality of the obtained optimal transition probabilities over probabilities obtained from other algorithms we have compared the transition probabilities obtained from Metropolis-Hasting algorithm with the optimal transition probabilities in a per step manner over the semi symmetric $K$-PPDR network depicted in Fig.2.

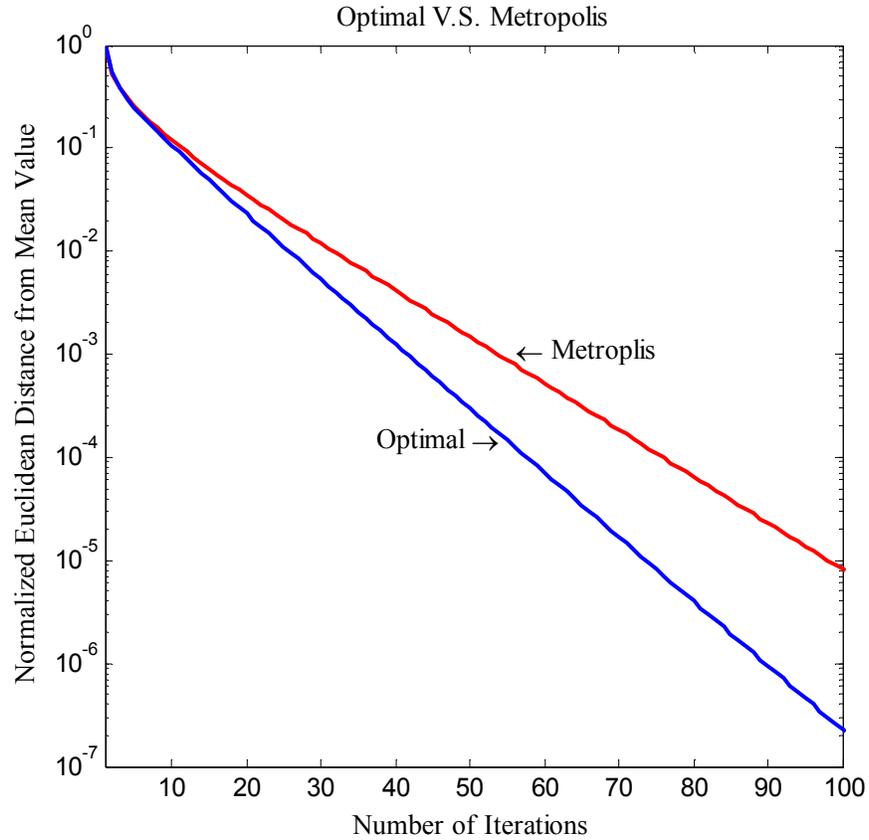

Fig.6. Normalized Euclidean Distance of vector of node values from the stationary distribution in terms of number of iterations.

In Fig.6. Normalized Euclidean distance of vector of node values from the stationary distribution in terms of number of iterations for optimal transition probabilities given in (5) and transition probabilities obtained from Metropolis-Hasting algorithm over the semi symmetric $K$-PPDR network depicted in Fig.2. is presented. As it is obvious from Fig.6. optimal transition probabilities have a better performance than the transition probabilities obtained from Metropolis-Hasting Algorithm. Also it should be mentioned that the results depicted in Fig. 6. are in logarithmic scale and generated based on 10000 trials (a different random initial node values is generated for each trial).

In networks with time invariant topologies optimal transition probabilities over a portion of edges depend on global topology of network while for the rest of the edges, their optimal transition probabilities only depend on local degree of their two incident nodes, thus the edges of a network can be divided into two disjoint groups, namely global and local edges, where the optimal transition probabilities over global edges unlike the optimal transition probabilities over local edges depend on global topology of network.

Comparing the optimal transition probabilities obtained for symmetric $K$-PPDR network and the optimal transition probabilities obtained for semi symmetric $K$-PPDR network we can conclude that the optimal transition probabilities for the edges of symmetric $K$-PPDR network can be determined by local degree of their two incident nodes whereas in a semi symmetric $K$-PPDR network for determining the optimal transition probabilities over edges connected to the nodes with strait edges more knowledge about the global topology of network is required, therefore in a semi symmetric $K$-PPDR network only the optimal transition probabilities on edges which their incident nodes are fully connected to all nodes of their neighboring sets can be considered as local edges and the rest of edges are global edges. In both cycle $K$-PPDR network and semi cycle $K$-PPDR network determining optimal transition probabilities over edges requires more knowledge about the global topology of network, thus all of their edges are global edges.

## V. Proofs of Main Results

In this section solution of FMMC problem and determination of optimal transition probabilities for symmetric $K$-PPDR network introduced in section III is presented.

Here we consider a symmetric $K$-PPDR network with $n$ nodes at each set and the undirected associated connectivity graph $\mathcal{G} = (\mathcal{V}, \mathcal{E})$. We denote the set of nodes on $i$-th set of symmetric $K$-PPDR graph by $(i, \mu)$ where $i$ and $\mu$ vary from 1 to $K$ and 1 to $n$ respectively.

Automorphism of symmetric $K$-PPDR graph is $S_n$ permutation of nodes of same set, hence according to subsection II-B it has $K$ class of edge orbits and it suffices to consider just $K-1$ transition probabilities $p_1, p_2, \ldots, p_{K-1}$ and consequently the transition probability matrix for the network can be defined as

$$P_{(i,\mu),(j,\rho)} = \begin{cases} 1 - np_1 & \text{for } i = j = 1, \quad \mu = \rho = 1, \ldots, n, \\ 1 - np_{i-1} - np_i & \text{for } i = j = 2, \ldots, K-1, \ \mu = \rho = 1, \ldots, n, \\ 1 - np_{K-1} & \text{for } i = j = K, \ \mu = \rho = 1, \ldots, n \\ p_i & \text{for } i = j - 1 = 1, \ldots, K-1, \ \mu, \rho = 1, \ldots, n \\ p_i & \text{for } i = j + 1 = 2, \ldots, K, \ \mu, \rho = 1, \ldots, n \end{cases}$$

We associate with the node $(i, \mu)$ the $|\mathcal{V}| \times 1$ column vector $e_{i,\mu} = e_i \otimes e_\mu$ for $\{i, \mu\} = \{i = 1, \ldots, K, \ \mu = 1, \ldots, n\}$ where $e_i$ and $e_\mu$ are $K \times 1$ and $n \times 1$ column vectors with one in the $i$-th and $\mu$-th position respectively and zero elsewhere. Introducing the new basis

$$\varphi_{i,q} = \frac{1}{\sqrt{n}} \sum_{m=0}^{n-1} \omega^{mq} e_{i,(m+1)} \quad \text{for} \quad q = 0, \ldots, n-1$$

with $\omega = e^{j\frac{2\pi}{n}}$, the transition probability matrix $P$ for Symmetric $K$-PPDR network in the new basis takes the block diagonal form with diagonal blocks $P_1, P_2, \ldots, P_n$ where $P_1$ is defined as:

$$P_1 = \begin{bmatrix} 1 - np_1 & np_1 & 0 & & & \\ np_1 & 1 - np_1 - np_2 & np_2 & 0 & & \\ 0 & np_2 & 1 - np_2 - np_3 & \ddots & 0 & \\ & 0 & & \ddots & \ddots & np_{K-1} \\ & & & & np_{K-1} & 1 - np_{K-1} \end{bmatrix} \qquad (7\text{-a})$$

and $P_i$ for $i = 2, \ldots, n$ is:

$$P_2 = \cdots = P_n = \begin{bmatrix} 1-np_1 & 0 & & & & \\ 0 & 1-np_2-np_3 & 0 & & & \\ & 0 & 1-np_3-np_4 & \ddots & & \\ & & \ddots & \ddots & & 0 \\ & & & & 0 & 1-np_K \end{bmatrix} \quad (7\text{-b})$$

We continue by assuming that both $\lambda_2(P)$ and $\lambda_{nK}(P)$ are amongst the eigenvalues of $P_1$ and the diagonal entries of $P_2$ which are its eigenvalues as well does not make any restriction on the obtained *SLEM* of network. After obtaining the SLEM of network we will show that the assumption still holds. Thus Based on subsection II-A, one can express FDC problem for symmetric $K$-PPDR network in the form of semidefinite programming as:

$$\begin{aligned} \min \quad & s \\ \text{s.t.} \quad & -sI \leq P_1 - \mathbf{1}\mathbf{1}^T \leq sI \\ & P \geq 0 \end{aligned} \quad (8)$$

where $\mathbf{1}$ is the column vector of all one, which is eigenvector of $W_1$ corresponding to the eigenvalue one. We relax the constraint on positivity of elements of transition probability matrix ($P \geq 0$) in (8) and continue the solution, after obtaining the optimal transition probabilities we will show that all elements of transition probability matrix are positive. Thus (8) reduces to following:

$$\begin{aligned} \min \quad & s \\ \text{s.t.} \quad & -sI \leq P_1 - \mathbf{1}\mathbf{1}^T \leq sI \end{aligned} \quad (9)$$

The matrix $P_1$ can be written as

$$P_1 = I_K - \sum_{i=1}^{K-1} p_i\, \boldsymbol{\alpha}_i \boldsymbol{\alpha}_i^T \quad (10)$$

where $\boldsymbol{\alpha}_i$ are $K \times 1$ column vectors defined as:

$$\boldsymbol{\alpha}_i(j) = \begin{cases} \sqrt{n} & j=i \\ -\sqrt{n} & j=i+1 \\ 0 & \text{Otherwise} \end{cases} \quad \text{for } i=1,\ldots,K-1$$

In order to formulate problem (9) in the form of standard semidefinite programming described in section II-C, we define $F_i, c_i$ and $x$ as below:

$$F_0 = \begin{bmatrix} -I_K + \mathbf{1}\mathbf{1}^T & 0 \\ 0 & I_K - \mathbf{1}\mathbf{1}^T \end{bmatrix}$$

$$F_i = \begin{bmatrix} \boldsymbol{\alpha}_i \boldsymbol{\alpha}_i^T & 0 \\ 0 & -\boldsymbol{\alpha}_i \boldsymbol{\alpha}_i^T \end{bmatrix} \quad \text{for } i=1,\ldots,K-1,$$

$$F_K = I_{2K}$$

$$c_i = 0, \ i = 1,\ldots K-1, \quad c_K = 1$$

$$x^T = [p_1, p_2, \ldots, p_{K-1}, s]$$

In the dual case we choose the dual variable $Z \geq 0$ as

$$Z = \begin{bmatrix} z_1 \\ z_2 \end{bmatrix} \cdot [z_1^T \quad z_2^T] \tag{11}$$

where $z_1$, and $z_2$ are $K \times 1$ column vectors. Obviously (11) choice of $Z$ implies that it is positive definite.

From the complementary slackness condition (3) we have

$$(sI - P_1 + \mathbf{11}^T)z_1 = 0 \tag{12-a}$$

$$(sI + P_1 - \mathbf{11}^T)z_2 = 0 \tag{12-b}$$

Multiplying both sides of equations (12) by $\mathbf{11}^T$ we have $s(\mathbf{11}^T z_1) = 0$ and $s(\mathbf{11}^T z_2) = 0$ which implies that

$$\mathbf{1}^T z_1 = 0 \tag{13-a}$$

$$\mathbf{1}^T z_2 = 0 \tag{13-b}$$

Using the constraints $Tr[F_i Z] = c_i$ we have

$$z_1^T z_1 + z_2^T z_2 = 1 \tag{14-a}$$

$$(\boldsymbol{\alpha}_i^T z_1)^2 = (\boldsymbol{\alpha}_i^T z_2)^2 \quad \text{for} \quad i = 1, \ldots, K-1 \tag{14-b}$$

To have the strong duality we set $c^T x + Tr[F_0 Z] = 0$, hence we have

$$z_1^T z_1 - z_2^T z_2 = s \tag{15}$$

Considering the linear independence of $\boldsymbol{\alpha}_i$ for $i = 1, \ldots, K-1$, we can expand $z_1$ and $z_2$ in terms of $\boldsymbol{\alpha}_i$ as

$$z_1 = \sum_{i=1}^{K-1} a_i \boldsymbol{\alpha}_i \tag{16-a}$$

$$z_2 = \sum_{i=1}^{K-1} a_i' \boldsymbol{\alpha}_i \tag{16-b}$$

with the coordinates $a_i$ and $a_i'$, $i = 1, \ldots, K-1$ to be determined.

Using (10) and the expansions (16), while considering (13), from comparing the coefficients of $\boldsymbol{\alpha}_i$ for $i = 1, \ldots, K-1$ in the slackness conditions (12), we have

$$(-s + 1)a_i = p_i \boldsymbol{\alpha}_i^T z_1, \tag{17-a}$$

$$(s + 1)a_i' = p_i \boldsymbol{\alpha}_i^T z_2, \tag{17-b}$$

where (17) holds for $i = 1, \ldots, K-1$. Considering (14-b), we obtain

$$(-s+1)^2 a_i^2 = (s+1)^2 a_i'^2,$$

for $i = 1, ..., K-1$, or equivalently

$$\frac{a_i^2}{a_j^2} = \frac{a_i'^2}{a_j'^2} \tag{18}$$

for $\forall i, j = [1, K-1]$ and for $\boldsymbol{\alpha}_i^T z_1$ and $\boldsymbol{\alpha}_i^T z_2$, we have

$$\boldsymbol{\alpha}_i^T z_1 = \sum_{j=1}^{K-1} a_j G_{i,j} \tag{19-a}$$

$$\boldsymbol{\alpha}_i^T z_2 = \sum_{j=1}^{K-1} a_j' G_{i,j} \tag{19-b}$$

where $G$ is the Gram matrices, defined as

$$G_{i,j} = \boldsymbol{\alpha}_i^T \boldsymbol{\alpha}_j$$

or equivalently

$$G = \begin{bmatrix} 2n & -n & 0 & \cdots & 0 \\ -n & 2n & -n & 0 & \vdots \\ 0 & -n & 2n & \ddots & 0 \\ \vdots & 0 & \ddots & \ddots & -n \\ 0 & \cdots & 0 & -n & 2n \end{bmatrix}$$

Substituting (19) in (17) we have

$$(-s + 1 - 2np_1)a_1 = -p_1 n a_2 \tag{20-a}$$

$$(-s + 1 - 2np_i)a_i = -p_i(na_{i-1} + na_{i+1}) \quad \text{for} \quad i = 2, ..., K-2 \tag{20-b}$$

$$(-s + 1 - 2np_{K-1})a_{K-1} = -p_{K-1} n a_{K-2} \tag{20-c}$$

and

$$(s + 1 - 2np_1)a_1' = -p_1 n a_2' \tag{21-a}$$

$$(s + 1 - 2np_i)a_i' = -p_i(na_{i-1}' + na_{i+1}') \quad \text{for} \quad i = 2, ..., K-2 \tag{21-b}$$

$$(s + 1 - 2np_{K-1})a_{K-1}' = -p_{K-1} n a_{K-2}' \tag{21-c}$$

Now we can determine the optimal transition probabilities in an inductive manner as follows:

In the first stage, from comparing equations (20-a) and (21-a) and considering the relation (18), we can conclude that

$$(-s + 1 - 2np_1)^2 = (s + 1 - 2np_1)^2$$

which results in $p_1 = 1/2n$ and $s = 0$, where the latter is not acceptable. Assuming $s = \cos(\theta)$ and substituting $p_1 = 1/2n$ in (20-a) and (21-a), we have

$$a_2 = \frac{\sin(2\theta)}{\sin(\theta)} a_1$$

$$a_2' = \frac{\sin(2(\pi - \theta))}{\sin(\pi - \theta)} a_1'$$

Continuing the above procedure inductively, up to $i - 1$ stages, and assuming

$$a_j = \frac{\sin(j\theta)}{\sin(\theta)} a_1, \qquad \forall j \leq i$$

and

$$a_j' = \frac{\sin(j(\pi - \theta))}{\sin(\pi - \theta)} a_1' \qquad \forall j \leq i$$

for the $i$-th stage, by comparing equations (20-b) and (21-c) we get the following equations

$$\left((-s + 1 - 2np_i)\frac{\sin(i\theta)}{\sin(\theta)} + np_i \frac{\sin((i-1)\theta)}{\sin(\theta)}\right) a_1 = -np_i a_{i+1} \tag{22-a}$$

$$\left((s + 1 - 2np_i)\frac{\sin(i(\pi - \theta))}{\sin(\pi - \theta)} + np_i \frac{\sin((i-1)(\pi - \theta))}{\sin(\pi - \theta)}\right) a_1' = -np_i a_{i+1}' \tag{22-b}$$

and considering relation (18) we can conclude that

$$\left((-s + 1 - 2np_i)\sin(i\theta) + np_i \sin((i-1)\theta)\right)^2 = \left((s + 1 - 2np_i)\sin(i(\pi - \theta)) + np_i \sin((i-1)(\pi - \theta))\right)^2$$

which results in

$$p_i = \frac{1}{2n} \tag{23}$$

Substituting $p_i = 1/2n$ in (22), we have

$$a_{i+1} = \frac{\sin((i+1)\theta)}{\sin(\theta)} a_1 \tag{24-a}$$

$$a_{i+1}' = \frac{\sin((i+1)(\pi - \theta))}{\sin(\pi - \theta)} a_1' \tag{24-b}$$

where (23) and (24) hold true for $i = 1, \ldots, K - 2$ and in the last stage, from equations (20-c) and (21-c) and using relations (18) and (24), we can conclude that

$$p_{K-1} = \frac{1}{2n} \tag{25}$$

and by substituting (25) in (20-c) we obtain

$$\sin(K\theta) = 0$$

which in turn results in $\theta_i = i\pi/K$ for $i = 1, \ldots, K-1$, therefore *SLEM* equals largest root of $s$ in magnitude which is $cos(\pi/K)$. Also one should notice that necessary and sufficient conditions for the convergence of transition probability matrix are satisfied, since all roots of $s$ which are the eigenvalues of $P$ are strictly less than one in magnitude, and one is a simple eigenvalue of $P$ associated with the eigenvector **1**, where this happens due to positivity of optimal transition probabilities [18].

Finally the dual constraints (14) and strong duality condition (15) are satisfied for $s = \cos(\pi/K)$ and the optimal transition probabilities given in (23) and (25). Also $a_1$ and $a_1'$ determined as

$$a_1 = \frac{1 + \cos(\theta)}{1 - \cos(\theta)} \times \frac{1}{n} \times \sin^2\theta$$

$$a_1' = \frac{1 - \cos(\theta)}{1 + \cos(\theta)} \times \frac{1}{n} \times \sin^2\theta$$

By considering the optimal transition probabilities (23) and (25) we can conclude that all elements of transition probability matrix are nonnegative and the elements of matrix $P_2$ given in (7-b) which are its eigenvalues as well, equal zero and $1/2$ which are both smaller than $\cos(\pi/K)$, for $\forall K \geq 3$ and in the special case of $K = 2$ by continuing the same procedure and assuming that $\lambda_2$ is amongst the eigenvalues of $P_2$ for optimal transition probabilities and *SLEM* of network we have

$$p_1 = \frac{2}{3n}$$

$$SLEM = \frac{1}{3}$$

## VI. CONCLUSION

Finding optimal transition probabilities for the problem of Fastest mixing Markov chain on networks with different topologies has been an active area of research for a number of years but most of the methods proposed so far usually avoid the direct computation of optimal transition probabilities and deal with the Fastest mixing Markov chain problem by numerical convex optimization methods.

Here in this work, we have solved Fastest mixing Markov chain problem for four particular types of $K$-partite networks by means of stratification and semidefinite programming. Our approach is based on fulfilling the slackness conditions, where the optimal transition probabilities are obtained by inductive comparing of the characteristic polynomials initiated by slackness conditions. The simulation results confirm that the Markov chain with optimal transition probabilities mixes substantially faster than the transition probabilities obtained from Metropolis-Hasting method and the obtained results proves that a symmetric $K$-PPDR network and its equivalent semi symmetric $K$-PPDR network have the same *SLEM* despite the fact that semi symmetric $K$-PPDR network has less edges than its equivalent symmetric $K$-PPDR network and at the same time symmetric *K*-PPDR network has better mixing rate per step than its equivalent semi symmetric *K*-PPDR network at first few iterations. The same results are true for cycle *K*-PPDR and semi cycle *K*-PPDR networks. We believe that the method used in this paper is powerful and lucid enough to be extended to networks with more general topologies which is the object of our future investigations.

**REFFERENCES**